\documentclass[11pt,preprint]{aastex}

\newcommand{\VEC}[1]{\mathbf{#1}}
\newcommand{\Mesz}{M\'esz\'aros}

\newcommand{\vrel}{v_{\mathrm{rel}}}
\newcommand{\vth}{v_{\mathrm{th}}}
\newcommand{\rIso}{r_{\mathrm{iso}}}
\newcommand{\Ppar}{P_{\|}}
\newcommand{\Pperp}{P_{\perp}}

\newcommand{\Tmntot}{T^{\mu\nu}_{\mathrm{tot}}}
\newcommand{\Tmntots}{T^{\mu\nu}_{\mathrm{tot,s}}}
\newcommand{\Tmntotals}{T_{\mathrm{tot,s}}}

\newcommand{\Tmnfluid}{T^{\mu\nu}_{\mathrm{fluid}}}
\newcommand{\TmnfluidB}{T_{\mathrm{fluid,B}}}
\newcommand{\Tmnfluidn}{T_{\mathrm{fluid,n}}}
\newcommand{\Tmnfluids}{T_{\mathrm{fluid,s}}}
\newcommand{\TmnEM}{T^{\mu\nu}_{\mathrm{EM}}}
\newcommand{\TmnEMn}{T_{\mathrm{EM,n}}}
\newcommand{\TmnEMs}{T_{\mathrm{EM,s}}}
\newcommand{\PmnfluidB}{\mathcal{P}_{\mathrm{fluid,B}}}
\newcommand{\FluxFluidEn}{F^{\mathrm{fluid}}_{en}}
\newcommand{\FluxFluidPx}{F^{\mathrm{fluid}}_{px}}
\newcommand{\FluxFluidPz}{F^{\mathrm{fluid}}_{pz}}
\newcommand{\FluxEmEn}{F^{\mathrm{EM}}_{en}}
\newcommand{\FluxEmPx}{F^{\mathrm{EM}}_{px}}
\newcommand{\FluxEmPz}{F^{\mathrm{EM}}_{pz}}
\newcommand{\TbnTwo}{\Theta_\mathrm{B2}}
\newcommand{\TBnTwo}{\Theta_\mathrm{B2}}
\newcommand{\TbnZ}{\Theta_\mathrm{B0}}
\newcommand{\TuTwo}{\Theta_\mathrm{u2}}

\newcommand\too{\, \rightarrow \, }

\newcommand{\gamrel}{\gamma_{\mathrm{rel}}}
\newcommand{\betarel}{\beta_{\mathrm{rel}}}

\newcommand{\rel}{relativistic}
\newcommand{\nonrel}{nonrelativistic}
\newcommand{\ultrarel}{ultrarelativistic}
\newcommand{\transrel}{trans-relativistic}

\newcount\Inum
\newcount\IInum
\newcount\IIInum
\newcount\IVnum
\def\I{\global\multiply\IInum by 0 \global\multiply\IIInum by 0
            \global\multiply\IVnum by 0 \global\advance \Inum by 1
            {\the\Inum. }}
\def\II{\global\multiply\IIInum by 0\global\multiply\IVnum by 0
       \global\advance \IInum by 1 {\the\Inum.\the\IInum. }}
\def\III{\global\multiply\IVnum by 0\global\advance \IIInum by 1
            {\the\Inum.\the\IInum.\the\IIInum. }}
\def\IV{\global\advance \IVnum by 1
            {\the\IVnum. }}
\def\bfrac#1#2{\hbox{${{\displaystyle#1 \vphantom{(} }\over{
   \displaystyle #2 \vphantom{(} }}$}}                
\newcount\listnorom
\newcommand\listromanDE{\global\advance \listnorom by 1 
{\lowercase\expandafter{(\romannumeral\listnorom)} }}
\newcommand\newlistroman{\listnorom=0}
\newcommand\TBn{\Theta_{\mathrm{B}}}
\newcommand\TBnZ{\Theta_{\mathrm{B0}}}

\newcommand\itt{ }
\newcommand\bff{ }

\newcommand\TP{test-particle}

\newcommand\Vsk{u_0}

\newcommand\iec{i.e.,}
\newcommand\egc{e.g.,}

\newcommand\alf{Alfv\'en}
\newcommand\Alf{Alfv\'en}

\newcommand{\MsonicZ}{M_\mathrm{S}}
\newcommand{\Malf}{M_\mathrm{A}}
\newcommand{\MalfZ}{M_\mathrm{A}}

\shorttitle{}
\shortauthors{}

\begin{document}     
\title{Magnetohydrodynamic Jump Conditions for Oblique Relativistic Shocks\\
with Gyrotropic Pressure}

\author{
Glen P. Double,\altaffilmark{1}
Matthew G. Baring,\altaffilmark{2}
Frank C. Jones,\altaffilmark{3}
and Donald C. Ellison\altaffilmark{1}
}

\altaffiltext{1}{Department of Physics, North Carolina State
 University, Box 8202, Raleigh, NC 27695,
\hbox{gpdouble@unity.ncsu.edu;}  don\_ellison@ncsu.edu}
\altaffiltext{2}{Department of Physics and Astronomy, Rice University,
MS 108, Houston, Texas 77251-1892, \hbox{baring@rice.edu}}
\altaffiltext{3}{Laboratory for High Energy Astrophysics, NASA Goddard
Space Flight Center, Greenbelt, MD 20771, \hbox{frank.c.jones@gsfc.nasa.gov}}

\slugcomment{Submitted to {\it Ap. J}, April 2003}

\begin{abstract}
Shock jump conditions, i.e., the specification of the downstream
parameters of the gas in terms of the upstream parameters, are obtained
for steady-state, plane shocks with oblique magnetic fields and
arbitrary flow speeds.  This is done by combining the continuity of
particle number flux and the electromagnetic boundary conditions at the
shock with the magnetohydrodynamic conservation laws derived from the
stress-energy tensor.  For 
\ultrarel\ and \nonrel\ shocks,
the
jump conditions may be solved analytically.
For mildly \rel\ shocks, analytic solutions are obtained for
isotropic pressure using an approximation for the adiabatic index that
is valid in high sonic Mach number cases.  Examples assuming isotropic
pressure illustrate how the shock compression ratio depends on the
shock speed and obliquity.  In the more general case of gyrotropic
pressure, the jump conditions cannot be solved analytically without
additional assumptions, and the effects of gyrotropic pressure are
investigated by parameterizing the distribution of pressure parallel
and perpendicular to the magnetic field.  Our numerical solutions
reveal that relatively small departures from isotropy (\egc\ $\sim
20$\%) produce significant changes in the shock compression ratio, $r$,
at all shock Lorentz factors, including \ultrarel\ ones, where an
analytic solution with gyrotropic
pressure is obtained.  In
particular, either dynamically important fields or significant
pressure anisotropies can incur marked departures from the canonical
gas dynamic value of $r=3$ for a shocked ultrarelativistic flow and
this may impact models of particle acceleration in gamma-ray bursts and
other environments where \rel\ shocks are inferred.  The jump
conditions presented apply directly to test-particle acceleration, and
will facilitate future self-consistent numerical modeling of particle
acceleration at oblique, \rel\ shocks; such models include the
modification of the fluid velocity profile due to the contribution of
energetic particles to the momentum and energy fluxes.
\end{abstract}

\newpage

\section{INTRODUCTION}
Collisionless shocks are pervasive throughout space and are regularly
associated with objects as diverse as stellar winds, supernova
remnants, galactic and extra-galactic radio jets, and accretion onto
compact objects.  Relativistic shocks, where the shock speed is close
to the speed of light, may be generated by the most energetic events;
for example, pulsar winds, blastwaves in quasars and active galactic
nuclei \citep[\egc][]{BM77}, and in gamma-ray bursts
\citep[\egc][]{Piran99}.  They may naturally emerge as the evolved
products of Poynting flux-driven or matter-dominated outflows in the
vicinity of compact objects such as neutron stars and black holes.  As
\rel\ shocks propagate through space, magnetic fields upstream from the
shock, at even small angles with respect to the shock normal, are
strongly modified by the Lorentz transformation to the downstream
frame.  The downstream magnetic fields are both increased and tilted
toward the plane of the shock and can have large angles with respect to
the shock normal.  Hence, most relativistic shocks can be expected to
see highly oblique magnetic fields and oblique magnetohydrodynamic
(MHD) jump conditions are required to describe them.

Relativistic shock jump conditions have been presented in a variety of
ways over the years.  The standard technique for deriving the
equations is to set the divergence of the stress-energy tensor equal
to zero on a thin volume enclosing the shock plane and use Gauss's
theorem to generate the jump conditions across the shock.  For
example, \citet{Taub48} developed the relativistic form of the
Rankine-Hugoniot relations, using the stress-energy tensor with
velocity expressed in terms of the Maxwell-Boltzmann distribution
function for a simple gas.  \citet{dHT50} presented a relativistic MHD
treatment of shocks in various orientations and a treatment of oblique
shocks for the nonrelativistic case, eliminating the electric field by
transforming to a frame where the flow velocity is parallel to the
magnetic field vector (now called the de Hoffmann-Teller frame).
\citet{Peacock81}, following \citet{LL59}, presented jump conditions
without electromagnetic fields, and \citet{BM76}, also using the
approach of \citet{LL59} and \citet{Taub48}, developed a concise set
of jump conditions for a simple gas using scalar pressure.
\citet{WZM87} provided a review of relativistic MHD shocks in ideal,
perfectly conducting plasmas, and in particular the treatment by
\citet{Lich67,Lich70}, which used this approach to develop the
relativistic analog of Cabannes' shock polar \citep{Cab70}, whose
origins also lie in \citet{LL59}.  \citet{KW88} developed hydrodynamic
equations using a pressure tensor, and \citet{AC88} developed
relativistic shock equations for MHD jets using scalar pressure and
magnetic fields with components $B_z$ and $B_{\phi}$ (a parallel field
with a twist).  \citet{BH91} derived MHD jump conditions using the
stress-energy tensor with isotropic pressure and the Maxwell field
tensor. By using a Lorentz transformation to the de Hoffman-Teller
frame, they restricted shock speeds, $\Vsk$, to $\Vsk/c < \cos{\TBn}$,
where $\TBn$ is the angle between the shock normal and the upstream
magnetic field and $c$ is the speed of light; hence, this approach may
only be used for mildly relativistic applications.

All of these approaches assumed that particles encountered by the
shock did not affect the shock structure, i.e., shocked particles were
treated as test particles.  Moreover, except for \citet{KW88}, they
confined their analyses to cases of isotropic pressures, a restriction
that is appropriate to thermal particles or very energetic ones
subject to the diffusion approximation in the vicinity of
non-relativistic shocks.  The assumption of pressure isotropy must be
relaxed when considering the hydrodynamics of relativistic shocks,
since their inherent nature imposes anisotropy on the ion and electron
distributions: this is due to the difficulty particles have streaming
against relativistic flows.  The computed particle anisotropies for
relativistic shocks are considerable in the shock layer
\cite[\egc][]{BedOstrow98,KGGA2000}, persisting up to arbitrarily high
ultrarelativistic energies.  The particles eventually relax to
isotropy in the fluid frame far downstream, on length scales
comparable to diffusive ones.  However, only a small minority of
accelerated particles achieve isotropy in the upstream fluid frame of
relativistic shocks, due to the rapid convection to the downstream
side of the flow discontinuity.  These isotropized particles are
present only when they manage to diffuse more than a diffusive mean
free path, $\lambda$, upstream of the shock; their contribution to the
flow dynamics is therefore dominated by that of the anisotropic
particles within a distance $\lambda$ of the shock.
For non-linear 
particle
acceleration,
where the
non-thermal ion or electron populations possess a sizable fraction of
the total energy or momentum fluxes, 
%
calculating pressure anisotropy will be critical for determining the
conservation of these fluxes through the shock transition
\citep[\egc][]{EBJ96}. The jump conditions we develop here can serve
as a guide to self-consistent solutions of non-linear relativistic shock
acceleration problems \citep{ED2002}.
%

\newlistroman

Here, we extend previous work by deriving a set of fully relativistic
MHD jump conditions with gyrotropic pressure and oblique magnetic
fields.  We adopt the gyrotropic case as a specialized generalization
because it 
\listromanDE
is exactly realized in plane-parallel shocks and is a
good approximation for oblique shocks where the flow deflection is
small, \iec\ the field plays a passive role (generally high Alfv\'enic
Mach number cases), and 
\listromanDE permits a comparatively simple expression of the
Rankine-Hugoniot jump conditions.  The sonic Mach number,
$\MsonicZ$, used throughout our paper, refers to the shock speed
compared to the fast-mode magnetosonic wave speed as described in
\citet{KD99}, i.e., $\MsonicZ \equiv \sqrt{3\rho_0 u_0^2/(5P_0)}$,
where $\rho_0$ is the upstream mass density and $P_0$ is the upstream
pressure.  The \Alf\ Mach number we refer to here is defined as
$\MalfZ \equiv \sqrt{4 \pi \rho_0 u_0^2}/B_0$ ($B_0$ is the upstream
magnetic field), regardless of the shock Lorentz factor, $\gamma_0 =
[1 - (u_0/c)^2]^{-1/2}$; i.e., we use these definitions applicable to
non-relativistic flows as parameters for the depiction of our results
at all $\gamma_0$.

Our results are not
restricted to the de Hoffmann-Teller frame and apply for arbitrary
shock speeds and arbitrary shock obliquities.  We solve these equations
and determine the downstream state of the gas in terms of the upstream
state first for the special case of isotropic pressure, and then, by
parameterizing the ratio of pressures parallel and perpendicular to the
magnetic field, for cases of gyrotropic pressure.

A principal result of this analysis is that {\it either} dynamically
important magnetic fields or significant pressure anisotropies produce
marked departures from the canonical value \citep[\egc][]{BM76, KD99}
of $r=3$ for the shock compression ratio in an ultrarelativistic
fluid.  The magnetic weakening of ultrarelativistic perpendicular
shocks came to prominence in the work of \citet{KC84} on the
interaction of the Crab pulsar's wind with its environment.  The
similarity of such consequences of fluid anisotropy and low Alfv\'enic
Mach number fields, which are also pervasive for trans-relativistic
and non-relativistic shocks, has its origin in the similar nature of
the plasma and electromagnetic contributions to the spatial components
of the stress-energy tensor.
This result may have important implications for the application of
first-order Fermi 
shock acceleration theory to gamma-ray bursts
and jets in active galaxies.

In this work, we concentrate on using analytic methods for determining
the fluid and electromagnetic characteristics of the shock and do not
explicitly include first-order Fermi particle acceleration. Future
work will combine these results with Monte Carlo techniques
\cite[\egc][]{EBJ96,ED2002} that will allow the modeling of Fermi
acceleration of particles, including the modification of the shock
structure resulting from the backreaction of energetic particles on
the upstream flow at all pertinent length scales. The jump conditions
we present here, however, apply directly to \TP\ Fermi acceleration in
shocks with arbitrary speed and obliquity.

\section{DERIVATION OF MHD JUMP CONDITIONS}
\subsection{Steady-State, Planar Shock}
 \label{sec:planarshock}
\begin{figure}[!hbtp]              
\epsscale{0.65} \plotone{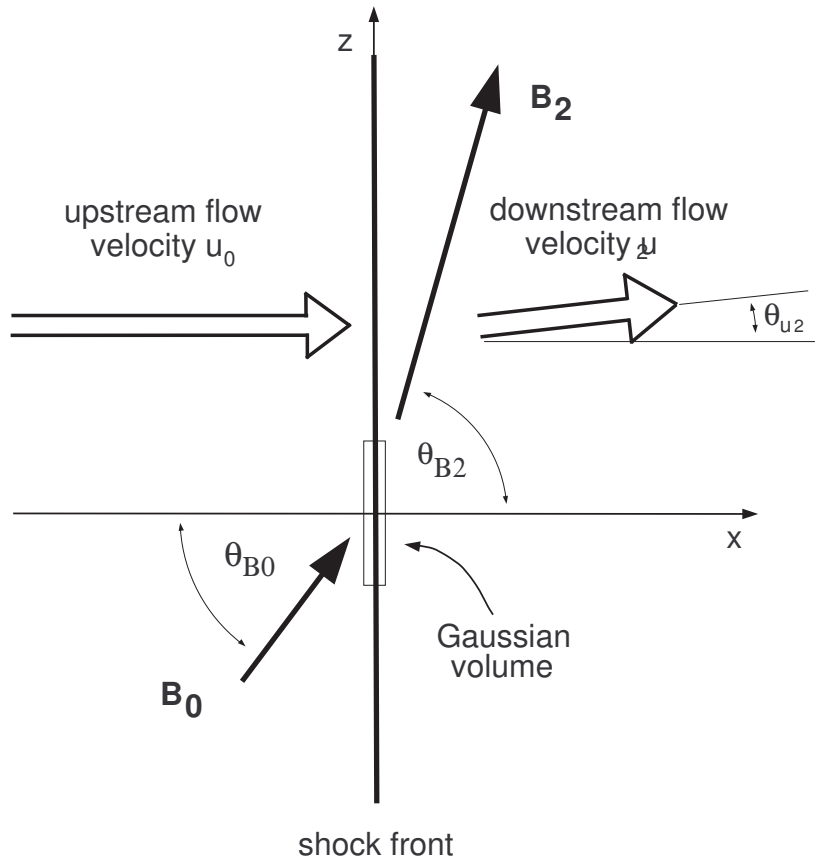} 
\figcaption{Schematic diagram of a plane shock, viewed from the shock
frame, showing oblique magnetic fields and a Gaussian volume over
which the divergence of the stress-energy tensor is integrated.  The
angles and field strengths are measured in the local plasma frame and,
in all of our examples, we take the upstream flow to be parallel to
the shock normal.  The $y$-axis is directed into the page.
\label{fig:sk_box}
}
\end{figure}

Using a Cartesian coordinate system with the $+x$-axis pointing
towards the downstream direction, we consider an infinite,
steady-state, plane shock traveling to the left at a speed $\Vsk$ with
its velocity vector parallel to the normal of the plane of the shock
as shown in Figure \ref{fig:sk_box}. The upstream fluid consists of a
thin, \nonrel\ plasma of protons and electrons in thermal equilibrium
with $\mathcal{T}_{p0} = \mathcal{T}_{e0}$, where $\mathcal{T}_{p0}$
($\mathcal{T}_{e0}$) is the unshocked proton (electron) temperature.
A uniform magnetic field, $B_0$, makes an angle $\TBnZ$ with respect
to the $x$-axis as seen from the upstream plasma frame.  We keep the
field weak enough to insure high \alf\ Mach numbers (\iec\ $\Malf
\gtrsim 2.5$) and thus to insure that the magnetic turbulence
responsible for scattering the particles is frozen into the plasma.
The $xyz$ coordinate system is oriented such that there are only two
components of magnetic field, $B_{x0}$ and $B_{z0}$, in the upstream
frame.  The field will remain co-planar in the downstream frame and
the downstream flow speed will be confined to the $x$-$z$ plane as
well \cite[e.g.,][]{JE87}.

In the shock frame, the upstream flow is in the $+x$-direction and is
described by the normalized 
four-velocity: $\beta^{\nu}_0 =
\gamma_0(1,\beta_0^x,0,0)$.  The downstream (i.e., shocked) flow
four-velocity 
is $\beta^{\nu}_2 = \gamma_2(1, \beta_2^x, 0,
\beta_2^z)$, where the subscript 0 (2) refers, here and elsewhere, to
upstream (downstream) quantities, {\boldmath $\beta$} $ =
\VEC{u}/c$, and $\gamma = (1 - \beta^2)^{-1/2}$ is the corresponding
Lorentz factor associated with the magnitude of the flow three-velocity
$\VEC{u}$.  Note that these subscript conventions follow those used in
\cite{EBB00}, where the subscript 1 is reserved for positions
infinitesimally upstream of the subshock discontinuity, admitting the
possibility of flow and field gradients upstream of the subshock.
Here the use of subscript 1 is redundant.  In addition, greek upper
and lower indices refer to spacetime components 0--3, while roman
indices refer to space components 1--3.

The set of equations connecting the upstream and downstream regions of
a shock consist of the continuity of particle number flux (for
conserved particles), momentum and energy flux conservation, plus
electromagnetic boundary conditions at the shock interface, and the
equation of state.  The various parameters that define the state of
the plasma, such as pressure and magnetic field, are determined in the
plasma frame and must be Lorentz transformed to the shock frame where
the jump conditions apply.  We assume that the electric field is zero
in the local plasma rest frame.  In general, the six jump conditions
plus the equation of state cannot be solved analytically because the
adiabatic index (i.e., the ratio of specific heats, whose value varies
smoothly between the \nonrel\ and \ultrarel\ limits) is a function of
the downstream plasma parameters, creating an inherently nonlinear
problem \citep[e.g.,][]{ER91}.  Even with the assumption of gyrotropic
pressure, there are more unknowns than there are equations; however,
with additional assumptions, approximate analytic solutions may be
obtained.

\subsection{Transformation Properties of the Stress-Energy Tensor}
%
The stress-energy tensor,
$T^{\mu\nu}$, describes the matter and electromagnetic momentum and
energy content of a medium at any given point in space-time.
Continuity across the shock is established by setting the appropriate
divergence (or covariant derivative $T^{\mu\nu}_{\ \ \ ;\nu}$) of the
stress-energy tensor equal to zero, namely for directions locally
normal to the shock.  Following standard expositions such as \cite{T34}
and \cite{Wein72}, the total stress-energy tensor, $\Tmntot$, is
expressed as the sum of fluid and electromagnetic parts, i.e.,
\begin{equation}
\label{ST3}
\Tmntot = \Tmnfluid + \TmnEM
\ .
\end{equation}
This is then integrated over a thin volume containing the shock plane
as shown in Figure \ref{fig:sk_box}. Application of Gauss's theorem
then yields the energy and momentum flux conditions across the plane
of the shock by using $T^{\mu\nu}n_{\nu} = 0$.  Accordingly,
$T^{0\nu}n_{\nu}$ yields the conservation of energy flux,
$T^{1\nu}n_{\nu}$ yields the $x$-contribution to momentum flux
conservation in the $x$-direction, $T^{3\nu}n_{\nu}$ yields the
$z$-contribution to momentum flux conservation in the $x$-direction,
and $n_{\nu} = (0,1,0,0)$ is the unit four-vector along the $x$-axis
in the reference frame of the shock.  The Einstein summation
convention is adopted here and elsewhere in this paper.

The components of the fluid and electromagnetic tensors are defined in
the local plasma frame and are subsequently Lorentz-transformed to the
shock frame where the flux conservation conditions apply, i.e.,
\begin{equation}
\label{ST4}
\Tmntots = 
\Lambda^{\mu}_{\alpha}
\left[ \Tmnfluidn^{\alpha\beta} + \TmnEMn^{\alpha\beta} \right]  
\Lambda^{\nu}_{\beta}
\ ,
\end{equation}
where the subscript $n$ ($s$) refers to the plasma (shock) frame
with the $x$-axis oriented normal to the shock in each case. 
Since the flow speeds in our model may have two space components, in
the $x$- and $z$-directions, the Lorentz transformation is
\begin{equation}
\label{LT0}
\Lambda^{\mu}_{\alpha} = 
\left( \begin{array}{cccc} 
\gamma & \gamma\beta_x & 0 & \gamma\beta_z\\[0.3cm]
\gamma\beta_x & 1 + \bfrac{\gamma-1}{\beta^2}\beta_x^2 & 
0 & \bfrac{\gamma-1}{\beta^2}\beta_x\beta_z\\[0.4cm]
0 & 0 & 1 & 0 \\[0.4cm]
\gamma\beta_z & \bfrac{\gamma-1}{\beta^2}\beta_z\beta_x & 
0 & 1 + \bfrac{\gamma-1}{\beta^2}\beta_z^2
\end{array}
\right) 
\ ,
\end{equation}
where the $\beta$ components and $\gamma$ are as previously defined in
Section~\ref{sec:planarshock}.  The system defined in
Figure~\ref{fig:sk_box} is invariant under translations in the
$y$-direction, so that conservation laws along the $y$-axis are
trivially satisfied.

\subsection{The Fluid Tensor and Equation of State}
The fluid tensor will be constrained to the gyrotropic case in the
local fluid frame; i.e., pressure can have one value parallel to the
magnetic field and a different value perpendicular to the magnetic
field (with symmetry about the magnetic field vector). This gives the
diagonal stress-energy tensor
\begin{equation}
\label{FTmag}
\TmnfluidB = 
\left(\begin{array}{cccc}
	e & 0 & 0 & 0\\
	0 & P_{\|} & 0 & 0\\
	0 & 0 & P_{\perp} & 0\\
	0 & 0 & 0 & P_{\perp}
\end{array} \right) 
\ ,
\end{equation}
where $e$ is the total energy density, $\Ppar$ ($\Pperp$) is the
pressure parallel (perpendicular) to the magnetic field, and the
subscript $B$ refers to the magnetic axis in the plasma frame.  We
obtain the fluid tensor in the $xyz$ plasma frame (labelled with
subscript $n$ referring to the shock normal),
$\Tmnfluidn$, with a rotation about the $y$-axis,
\begin{equation}
   \Tmnfluidn \;=\; \mathcal{R} \;\TmnfluidB\; \mathcal{R}^{-1}\ ,
\end{equation}
where
\begin{equation}
\label{rotator}
\mathcal{R} = 
\left( \begin{array}{cccc}
        1 & 0 & 0 & 0\\
        0 & \cos\TBn  & 0 & \sin\TBn \\
        0 & 0 & 1 & 0\\
        0 & -\sin\TBn & 0 & \cos\TBn
\end{array} \right) 
\ .
\end{equation}
The resulting tensor in the {\it{xyz}} plasma frame is
\begin{equation}
\label{FTrot}
\Tmnfluidn = 
 \left( \begin{array}{cccc}
 e & 0 & 0 & 0\\
 0 & P_{\|}\cos^2{\TBn} + P_{\perp}\sin^2{\TBn} & 0 &
 (P_{\perp}-P_{\|})\sin{\TBn}\cos{\TBn}\\
 0 & 0 & P_{\perp} & 0 \\
 0 & (P_{\perp}-P_{\|})\sin{\TBn}\cos{\TBn} & 0 & 
 P_{\perp}\cos^2{\TBn}+P_{\|}\sin^2{\TBn}
\end{array} \right) 
\ ,
\end{equation}
with the $T^{ij}$ space components corresponding to the 3-dimensional
pressure tensor presented by \cite{EBJ96} for \nonrel\ shocks, once a
typographical error is corrected.

Renaming the components of the fluid tensor as
\begin{equation}
\label{fluidT2}
\Tmnfluidn = 
\left( \begin{array}{cccc}
	e & 0 & 0 & 0 \\[0.1cm]
	0 & P_{xx} & 0 & P_{xz} \\[0.2cm]
	0 & 0 & P_{yy} & 0 \\[0.2cm]
	0 & P_{zx} & 0 & P_{zz}  
\end{array} \right) 
\ ,
\end{equation}
with $P_{xz} = P_{zx}$, we have the identities
\begin{equation}
\label{P21}
\Ppar = P_{xx} - P_{xz}\bfrac{\sin{\TBn}}{\cos{\TBn}}
\ ,
\end{equation}
\begin{equation}
\label{P22}
\Pperp = P_{xx} + P_{xz}\bfrac{\cos{\TBn}}{\sin{\TBn}}
\ ,
\end{equation}
and,
\begin{equation}
\label{Pzz}
P_{zz} = P_{xx} + 2P_{xz}\cot{(2\TBn)}
\ .
\end{equation}
%
The $T^{00}$ component,
$e$, is the total energy density in the rest or plasma frame. The other
components, $P_{ij}$, are defined by \citet{T34} as the ``absolute
stress'' components in the proper frame. $P_{ij}$ is the pressure
parallel to the $i$-axis exerted on a unit area normal to the
$j$-axis.  Hence, the diagonal components can be considered a pressure,
but the off-diagonal components are {\it{shear}} stresses.  The
isotropic scalar pressure, $P=Tr(\PmnfluidB )/3$, is a Lorentz
invariant.  The fluid tensor, in general, changes its appearance
significantly under a Lorentz transformation via the mixing of
components: for example, it picks up  momentum flux components in
reference frames moving with respect to the proper frame \citep{T34}.

Using the above expressions, we can derive an adiabatic equation of
state.  Starting with the spatial portion of the fluid tensor:
\begin{equation}
\label{PT01}
\PmnfluidB =
\left( \begin{array}{ccc}
	P_{\|} & 0 & 0\\
	0 & P_{\perp} & 0\\
	0 & 0 & P_{\perp}
\end{array} \right) 
\ ,
\end{equation}
the total energy density can be written as
\begin{equation}
\label{cons_nrgy}
e = \bfrac{Tr(\PmnfluidB )}{3(\Gamma - 1)} + \rho c^2
\ ,
\end{equation}
where $Tr(\PmnfluidB )$ is the trace of the pressure tensor,
$\Gamma$ is the adiabatic index, and $\rho c^2$ is the rest mass
energy density. Using equations (\ref{P21}) and (\ref{P22}),
\begin{equation}
\label{nr_eos2}
\bfrac{Tr(\PmnfluidB )}{3(\Gamma - 1)} = 
\bfrac{1}{3(\Gamma - 1)}(P_{\parallel} + 2P_{\perp}) =
\bfrac{1}{\Gamma - 1}\left[P_{xx} + \bfrac{P_{xz}}{3}
\left(2\cot{\TBn} - \tan{\TBn}\right)\right]
\ .
\end{equation}
In terms of the magnetic field, where $\tan\TBn = B_z/B_x$, the
adiabatic, gyrotropic equation of state becomes
\begin{equation}
\label{rel_eos}
e = 
\bfrac{1}{\Gamma - 1}
\left[P_{xx} + 
\bfrac{P_{xz}}{3}
\left(2\bfrac{B_x}{B_z} - \bfrac{B_z}{B_x}\right)\right]
+ \rho c^2
\ .
\end{equation}
While $\Gamma$ is well defined in the \nonrel\ and fully \rel\ limits,
in mildly \rel\ shocks $\Gamma$ depends on an unknown relation between
$e$ and $P$ which we will address in a later section.

\subsection{The Electromagnetic Tensor}
The electric and magnetic field components of the general electromagnetic 
tensor in the plasma frame are given by \citet{T34} as
\begin{equation}
\label{EMT}
\TmnEMn \; =\; 
\bfrac{1}{4\pi}\left( \begin{array}{cccc}
\bfrac{E^2+B^2}{2} & \left(\VEC{E}\times\VEC{B}\right)_x & 
\left(\VEC{E}\times\VEC{B}\right)_y & 
\left(\VEC{E}\times\VEC{B}\right)_z \\[0.2cm]
\left(\VEC{E}\times\VEC{B}\right)_x & Q_{xx} & Q_{xy} & Q_{xz} \\[0.3cm]
\left(\VEC{E}\times\VEC{B}\right)_y & Q_{yx} & Q_{yy} & Q_{yz} \\[0.3cm]
\left(\VEC{E}\times\VEC{B}\right)_z & Q_{zx} & Q_{zy} & Q_{zz}  
\end{array} \right) 
\ ,
\end{equation}
where the {\it{Q}}'s are the Maxwell stresses defined as
\begin{equation}
Q_{ii} = \bfrac{E^2 + B^2}{2} - E^2_i - B^2_i \quad
{\rm and} \quad
Q_{ij} = -(E_iE_j + B_iB_j)
\ .
\end{equation}
Note that here the suffix $n$ denotes an axis orientation along the
shock normal, as it does for the fluid tensor.  Since the electric
fields in the plasma frame are negligible, this simplifies to
\begin{equation}
 \label{EMT11}
\TmnEMn \; =\; 
\bfrac{1}{4\pi}\left( \begin{array}{cccc}
\bfrac{B^2}{2} & 0 & 0 & 0 \\ 
0 & \bfrac{B^2_z - B^2_x}{2} & 0 & -B_xB_z \\
0 & 0 & \bfrac{B^2}{2} & 0 \\
0 & -B_zB_x & 0 & \bfrac{B^2_x - B^2_z}{2} 
\end{array} \right) 
\ .
\end{equation}
Observe that this electromagnetic contribution to the stress-energy
tensor resembles the structure of that for the gyrotropic fluid in
equation~(\ref{fluidT2}), \iec\ the presence of laminar fields should
mimic anisotropic pressures in terms of their effect on the flow
dynamics.  A noticeable difference, however, is that while the
magnetic field can exhibit tension and can therefore generate negative
diagonal components for $T^{\mu\nu}$, depending on the orientation of
the field, the corresponding diagonal pressure components are always
positive definite.

\subsection{Flux Conservation Relations}
As discussed above, the energy and momentum conservation relations in
the shock frame can be derived by applying $T^{\mu\nu}n_{\nu} = 0$ to
equation (\ref{ST4}), individually on the Lorentz-transformed fluid
and electromagnetic tensors.
The conservation of energy flux derives from
\begin{equation}
\label{TE1}
\Tmntotals^{0\nu}\, n_{\nu} = 
\Tmnfluids^{0\nu}\, n_{\nu} +  
\TmnEMs^{0\nu}\, n_{\nu} = 0
\ .
\end{equation}
The fluid contribution to energy flux conservation is 
\begin{eqnarray}
\label{PE42}
\FluxFluidEn =
\Tmnfluids^{0\nu}\, n_{\nu} = 
\gamma^2 \beta_x (e + P_{xx}) -
\gamma (\gamma-1) \bfrac{\beta_x \beta_z^2}{\beta^2}(P_{xx}-P_{zz}) + 
\hspace{3cm} \nonumber \\
\gamma \left[(2\gamma - 1)\beta_x^2 + \beta_z^2\right]
\bfrac{\beta_z}{\beta^2}P_{xz} 
\ ,
\end{eqnarray}
while the electromagnetic contribution is 
\begin{equation}
\label{CE33}
\FluxEmEn = 
\TmnEMs^{0\nu}\, n_{\nu} = 
\bfrac{\gamma}{4\pi\beta^2}
\left[(\gamma -1)\beta_x\beta_z^2B_x^2 + 
(\gamma\beta_x^2 + \beta_z^2)\beta_xB_z^2 -
\{(2\gamma - 1)\beta_x^2 + \beta_z^2\}\beta_zB_xB_z\right]
\ .
\end{equation}

The conservation of momentum flux derives from
\begin{equation}
\label{TM1}
\Tmntotals^{i\nu}\, n_{\nu} = 
\Tmnfluids^{i\nu}\, n_{\nu} +  
\TmnEMs^{i\nu}\, n_{\nu} = 0
\ .
\end{equation}
The $x$-component of the transformed fluid tensor 
contributing to the conservation of momentum flux is
\begin{eqnarray}
\label{PM44}
\FluxFluidPx = 
\Tmnfluids^{1\nu}\, n_{\nu} = 
\gamma^2\beta_x^2(e + P_{xx}) + P_{xx} -
(\gamma-1)^2\bfrac{\beta_x^2\beta_z^2}{\beta^4}(P_{xx} - P_{zz}) + 
\hspace{2cm} \nonumber \\
2(\gamma-1)(\gamma\beta_x^2 + \beta_z^2)
\bfrac{\beta_x\beta_z}{\beta^4}P_{xz}  
\ ,
\end{eqnarray}
and the $z$-component is
\begin{eqnarray}
\label{PM46}
\FluxFluidPz = 
\Tmnfluids^{3\nu}\, n_{\nu} = 
\gamma^2\beta_x\beta_z(e + P_{xx}) - 
(\gamma-1)(\beta_x^2 + \gamma\beta_z^2)\bfrac{\beta_x\beta_z}{\beta^4}
(P_{xx} - P_{zz}) +
\hspace{1.5cm}  \nonumber \\
\left[\gamma + 2(\gamma-1)^2\bfrac{\beta_x^2\beta_z^2}{\beta^4}\right]P_{xz}
\ .
\end{eqnarray}

The $x$-component of the electromagnetic contribution to the
conservation of momentum flux is
\begin{eqnarray}
\label{CM32}
\FluxEmPx = 
\TmnEMs^{1\nu}\, n_{\nu} = 
\bfrac{\gamma^2}{8\pi}\beta_x^2B^2 +
\bfrac{1}{8\pi\beta^4}\left[(\gamma\beta_x^2 + \beta_z^2)^2 - 
(\gamma - 1)^2\beta_x^2\beta_z^2\right](B_z^2 - B_x^2) - 
\hspace{1cm} \nonumber \\
\bfrac{1}{2\pi}(\gamma-1)(\gamma\beta_x^2 + \beta_z^2)
\bfrac{\beta_x\beta_z}{\beta^4}
B_xB_z 
\ ,
\end{eqnarray}
and the $z$-component is
\begin{eqnarray}
\label{CM34}
\FluxEmPz =
\TmnEMs^{3\nu}\, n_{\nu} = 
\bfrac{\gamma^2}{8\pi}\beta_x\beta_zB^2 +
\bfrac{1}{8\pi}(\gamma-1)^2(\beta_x^2 - \beta_z^2)
\bfrac{\beta_x\beta_z}{\beta^4}(B_z^2 - B_x^2) - \hspace{2cm}
\nonumber \\
\bfrac{1}{2\pi}(\gamma - 1)^2\bfrac{\beta_x^2\beta_z^2}{\beta^4}B_xB_z -
\bfrac{\gamma}{4\pi}B_xB_z
\ .
\end{eqnarray}
Clearly, parallels between the various components of the fluid and
electromagnetic stress tensors can be drawn by inspection of these
forms for the fluxes.

In all cases where the \Alf\ Mach number is greater than a few,
the downstream flow velocity deviates only slightly from the shock
normal direction so $\beta_z \ll \beta_x$. This allows a first-order
approximation in $\beta_z$ and the above equations become:
\begin{eqnarray}
   \FluxFluidEn 
   & \approx &
   \gamma^2\beta_x(e + P_{xx}) + \gamma(2\gamma-1)\beta_zP_{xz}
   \ ,\nonumber\\[-5.5pt]
 \label{eq:enflux}\\[-5.5pt]
   \FluxEmEn
   & \approx &
   \bfrac{\gamma^2}{4\pi}\beta_xB_z^2 - 
   \bfrac{\gamma(2\gamma-1)}{4\pi}\beta_zB_xB_z\ ,\nonumber
\end{eqnarray}
for the energy flux contributions, and
\begin{eqnarray}
   \FluxFluidPx
   & \approx &
   \gamma^2\beta_x^2(e + P_{xx}) + P_{xx} + 
   2\gamma(\gamma-1)\bfrac{\beta_z}{\beta_x}P_{xz}
   \ ,\nonumber\\[-5.5pt]
 \label{eq:pxflux}\\[-5.5pt]
   \FluxEmPx
   & \approx &
   \bfrac{\gamma^2}{8\pi}\beta_x^2B^2 +
   \bfrac{\gamma^2}{8\pi}(B_z^2 - B_x^2) - 
   \bfrac{\gamma(\gamma-1)}{2\pi}\bfrac{\beta_z}{\beta_x}B_xB_z
   \ ,\nonumber
\end{eqnarray}
and
\begin{eqnarray}
   \FluxFluidPz
   & \approx & 
   \gamma^2\beta_x\beta_z(e + P_{xx}) + \gamma P_{xz} - 
   (\gamma-1)\bfrac{\beta_z}{\beta_x}(P_{xx} - P_{zz}) 
   \ ,\nonumber\\[-5.5pt]
 \label{eq:pzflux}\\[-5.5pt]
   \FluxEmPz
   & \approx & 
   \bfrac{\gamma^2}{8\pi}\beta_x\beta_zB^2 +
   \bfrac{1}{8\pi}(\gamma-1)^2\bfrac{\beta_z}{\beta_x}(B_z^2 - B_x^2) - 
   \bfrac{\gamma}{4\pi}B_xB_z \nonumber
\end{eqnarray}
for the $x-$ and $z-$components of momentum flux, respectively.  As we
show below, the approximation, $\beta_z \ll \beta_x$, becomes
progressively better as the shock Lorentz factor increases but, in
fact, equations~(\ref{eq:enflux}-\ref{eq:pzflux}) provide an excellent
approximation at {\it all Lorentz factors} for virtually the entire
parameter regime we have considered in this paper. Unless Mach numbers
less than a few and/or extreme anisotropies are considered,
equations~(\ref{eq:enflux}-\ref{eq:pzflux}) yield solutions within one
part in $10^4$ to those obtained with
equations~(\ref{PE42}-\ref{CM34}) and are much easier to solve.

\subsection{Jump Conditions}
The jump conditions consist of the energy and momentum flux
conservation relations, the particle flux continuity, and the boundary
conditions on the magnetic field.  The conservation of particle number
flux\footnote{We assume there is no pair creation nor annihilation.}
is
\begin{equation}
\label{numdens}
\bigg[\gamma n\beta_x\bigg]^2_0 = 0
\ ,
\end{equation}
where the brackets provide an abbreviation for
\begin{equation}
\gamma_2 n_2 \beta_{x2} -
\gamma_0 n_0 \beta_{x0} = 0
\ .
\end{equation}
This jump condition, as well as the ones that follow, are written in
the shock frame and, as always, the subscript 0 (2) refers to upstream
(downstream) quantities.  The remaining jump conditions are:
\begin{eqnarray}
   \bigg[\FluxFluidEn  + \FluxEmEn \bigg]^2_0 &=& 0 \ ,\nonumber\\[2pt]
   \bigg[ \FluxFluidPx + \FluxEmPx \bigg]^2_0 &=& 0 \ ,
 \label{eq:jump_cond}\\[2pt]
   \bigg[\FluxFluidPz + \FluxEmPz \bigg]^2_0 &=& 0 \ .\nonumber
\end{eqnarray}
Adding the steady-state conditions on the magnetic field, namely that
$\nabla \cdot \hbox{\bf B}=0$:
\begin{equation}
\label{s4Bnorm}
\bigg[B_x\bigg]^2_0 = 0
\ ,
\end{equation}
and also that $\nabla\times \hbox{\bf E} = 0$:
\begin{equation}
 \label{s4Etan2}
\bigg[\gamma(\beta_zB_x - \beta_xB_z)\bigg]^2_0 = 0
\ ,
\end{equation}
completes the set of six jump conditions.  In the limit of $(\gamma
-1)\ll 1$ and for isotropic pressures, the above expressions reproduce
the standard continuity conditions at non-relativistic shocks
\cite[\egc\ see p.~117 of][]{BS69}.

At this point there are eight unknown downstream quantities
($\beta_{x2}$, $\beta_{z2}$, $B_{x2}$, $B_{z2}$, $P_{xx2}$, $P_{xz2}$,
$e_2$, and $n_2$) and only six equations.  If isotropic pressure is
assumed, $P_{xx} = P_{zz} = P$ and $P_{xz} = 0$, leaving six equations
and seven unknowns.  To obtain a closed set of equations for isotropic
pressure, an assumed equation of state (e.g., equation~\ref{rel_eos})
is added to the analysis.  Successive elimination of variables then
generally leads to a 7th-order equation in the compression ratio
$r\equiv \beta_{x0}/\beta_{x2}$ with lengthy algebraic expressions for
its coefficients
\citep[e.g.,][]{WZM87,AC88}.  The forms of the coefficients depend
on the assumed equation of state and 
do not simplify easily.
Here we perform some of the simpler algebraic eliminations and then
use a Newton-Raphson technique to iteratively solve
two simultaneous equations.

In Section~\ref{sec:isotropicP} below, we derive an approximate
expression for the downstream adiabatic index, $\Gamma_2$, for cold
upstream plasmas.  For anisotropic cases, further microphysical
information is required, which is generally only accessible using
computer simulations; in the gyrotropic approximation, we parameterize
pressure anisotropy in Section~\ref{sec:gyrotropicP} to provide
insight into global characteristics of the jump conditions.

\section{RESULTS}

\subsection{Isotropic Pressure, High Sonic Mach Number Cases}
 \label{sec:isotropicP}
Oblique shock jump conditions cannot, in general, be solved
analytically even for isotropic pressure because the downstream
adiabatic index, $\Gamma_2$, depends on the total downstream energy
density and the components of the pressure tensor (or scalar
pressure), which are not known before the solution is obtained. The
problem is inherently nonlinear except in the nonrelativistic and
ultra-relativistic limits where $\Gamma_2 = 5/3$ and $4/3$,
respectively.  Furthermore, the gyrotropic pressure components are
determined by the physics of the model and do not easily lend
themselves to analytic interpretation, although \citet{KW88} provided
equations based on a power-law distribution in momentum for the
pressure tensor components in the special case of a parallel
relativistic shock with test particle first-order Fermi shock
acceleration.

\begin{figure}[!hbtp]              
\epsscale{0.85} \plotone{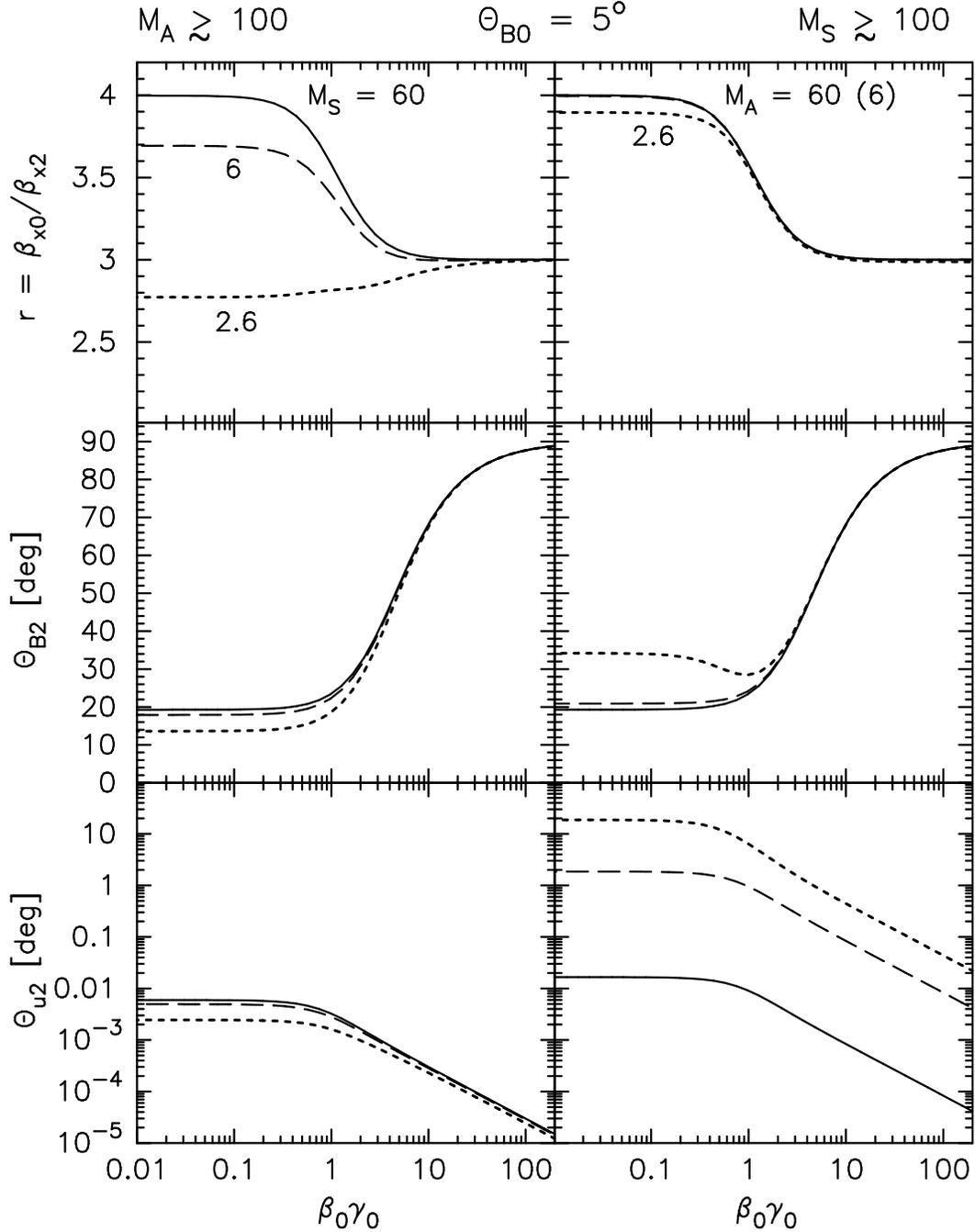} 
\figcaption{Compression ratio, $r$, $\TbnTwo$, and $\TuTwo$ versus
$\beta_0 \gamma_0$.  The three panels on the left have $\MalfZ
\gtrsim 100$ and the sonic Mach numbers as shown. The three right-hand
panels have $\MsonicZ \gtrsim 100$ with the \alf\ Mach numbers as
shown. In all cases, $\TbnZ=5^\circ$. Note that in the top right-hand
panel, the $\MalfZ=6$ and 60 curves are nearly identical.
\label{fig:T5_6}
}
\end{figure}

An excellent approximation can be obtained in the absence of efficient
particle acceleration if $\vth \ll u_0$, where $\vth$ is
the thermal speed of the unshocked plasma.  
In this case, which
corresponds to high sonic Mach numbers, upon scattering in the
downstream frame all particles have
\begin{equation}
\gamrel \simeq (1 - \betarel^2)^{-1/2}
\ ,
\end{equation}
where
\begin{equation}
\label{betarel_eq}
\betarel = \frac{\vrel}{c} = 
\frac{\beta_0 - \beta_2}{1-\beta_0\beta_2}
\end{equation}
is the relative $\beta$ between the converging plasma frames.
From kinetic theory, the isotropic pressure is
\begin{equation}
\label{eq:KE_P}
P = \frac{n}{3}<\VEC{p}\cdot\VEC{v}>
\ ,
\end{equation}
where $n$ is the particle number density, and $\VEC{p}$ and $\VEC{v}$
are the particle momentum and velocity, respectively.  Then, with our
approximation for particle velocity and using the isotropic version of
equation~(\ref{cons_nrgy}), 
\begin{equation}
\Gamma_2 = 
\frac{P}{e - \rho c^2} + 1 =
\frac{(1/3) p_{\mathrm{rel}} v_{\mathrm{rel}}}{(\gamrel - 1)m c^2} + 1 =
\frac{\gamrel \betarel^2}{3 (\gamrel - 1)} + 1
\ ,
\end{equation}
or,
\begin{equation}
\label{SpecHeat}
\Gamma_2 = \frac{4 \gamrel + 1}{3 \gamrel}
\ .
\end{equation}
This essentially kinematic approximation, which is operable only if
diffusive transport of particles from downstream to upstream
contributes insignificantly to the momentum and energy fluxes, permits
a direct numerical solution for isotropic pressure, arbitrary obliquity
(as long as the upstream Alfv\'enic Mach number, $\MalfZ$, is high),
and arbitrary flow speed \citep[see, for example,][for alternative
  forms for $\Gamma_2$]{Kirk88,Gallant2002}.  Note that equation
  (\ref{SpecHeat}) provides an upper limit to the adiabatic index
  because any particles accelerated by the shock would tend to raise
  the average Lorentz factor and cause the adiabatic index to
  decrease.

\begin{figure}[!hbtp]              
\epsscale{0.85} \plotone{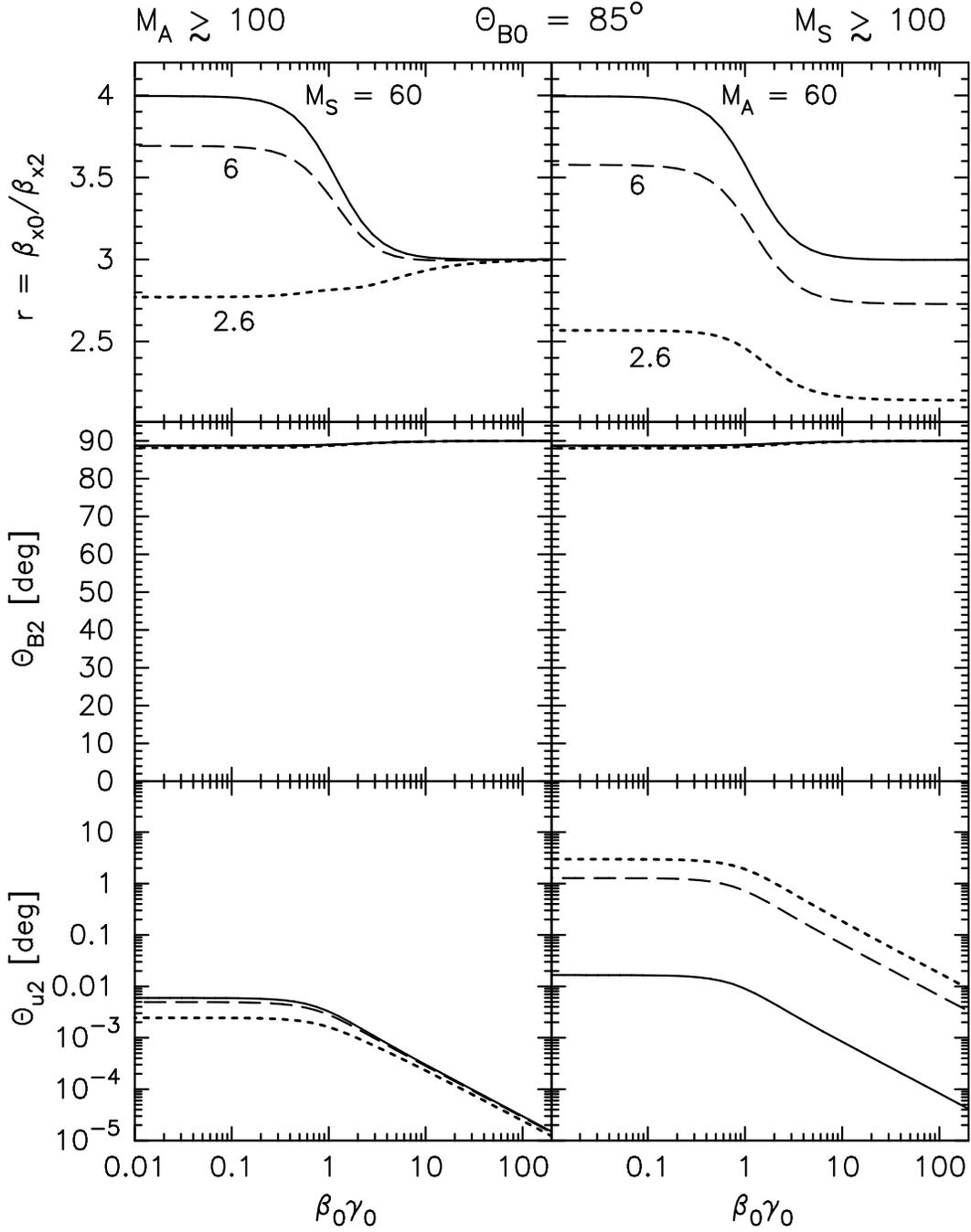} 
\figcaption{Same as in Figure~\ref{fig:T5_6} except all examples here
have $\TbnZ=85^\circ$. Note the drop in $r$, even at $\gamma_0 \gg 1$,
in the upper right-hand panel where anisotropic magnetic stresses
become important.
\label{fig:T85_6}
}
\end{figure}

In Figures~\ref{fig:T5_6} and \ref{fig:T85_6} we show results for the
compression ratio, $r$, and the downstream angles, $\TbnTwo$ and
$\TuTwo$, as a function of $\beta_0 \gamma_0$, for two extreme
upstream magnetic field angles, $\TbnZ$, and various sonic and \alf\
Mach numbers. 
The magnitude of the downstream field is
\begin{equation}
\label{eq:DSmag}
B_2 = [ B_{x0}^2 + \gamma_0^2 (r^2 - 1) B_{z0}^2 ]^{1/2}
\ .
\end{equation}
%
 
There are a number of important characteristics of these results.
First, the jump conditions map smoothly from fully \nonrel\ to
\ultrarel\ shock speeds and obtain the canonical values for the
compression ratio $r \equiv \beta_{x0}/\beta_{x2} = 4$ for high Mach
number, \nonrel\ shocks, and $r=3$ for high Mach number \ultrarel\
shocks.  For $\beta_0\gamma_0 \lesssim 1$ and regardless of the
obliquity, a low $\MsonicZ$ results in a weaker shock with smaller
$r$, as expected \citep[similar behavior is exhibited in Figure~1
of][]{AC88}. For $\beta_0\gamma_0 \gtrsim 10$, the sonic Mach number
$\MsonicZ$ has little influence on the results until it becomes very
low, i.e.  $\gamma_0 \MsonicZ^2\sim 1$.  Since our definition of
$\MsonicZ$ is inherently non-relativistic, perceptible changes to the
fluid dynamics arise only when the pressure $P_0$ becomes comparable
to the relativistic ram pressure $\gamma_0\beta_0^2\rho_0 c^2$; this
domain is exhibited in the parallel fluid shock jump conditions
explored by \citet{Taub48}.

Figure~\ref{fig:T5_6} with $\TbnZ=5^\circ$ and Figure~\ref{fig:T85_6}
with $\TbnZ=85^\circ$ illustrate an important characteristic of
\rel\ shocks.  For all upstream field obliquities other than $\TbnZ=0$,
the downstream magnetic field angle shifts towards $\TbnTwo=90^\circ$
as the shock Lorentz factor increases, indicating the importance of
addressing oblique fields when treating acceleration at highly \rel\
shocks.  
The transition criterion is directly obtainable from 
equation~(\ref{s4Etan2}) (since $\TuTwo$ is, in general, very small).
One quickly arrives at
$\beta_0 \gamma_0 \tan\TbnZ\gtrsim 1$ being the necessary condition to
render $\tan\TbnTwo > 1$, so that $\TbnTwo \too 90^\circ$.
Furthermore, in the left hand panels of Figures~\ref{fig:T5_6}
and~\ref{fig:T85_6}, the angle the downstream flow makes with the shock
normal, $\TuTwo$, is small at all $\gamma_0\beta_0$ (note the
logarithmic scale for $\TuTwo$), consistent with the assumption that
$\beta_z \ll \beta_x$.  This is a consequence of the passive magnetic
field corresponding to $\MalfZ\gg 1$.  Very different behavior is
exhibited when the field becomes dynamically important, as is evident
in the upper right-hand panels of Figures~\ref{fig:T5_6}
and~\ref{fig:T85_6}:  a low $\MalfZ$ (\iec\ high $B_0$) can drastically
lower $r$, even at \ultrarel\ speeds \cite[see][]{KC84}.

\begin{figure}[!hbtp]              
\epsscale{0.7} \plotone{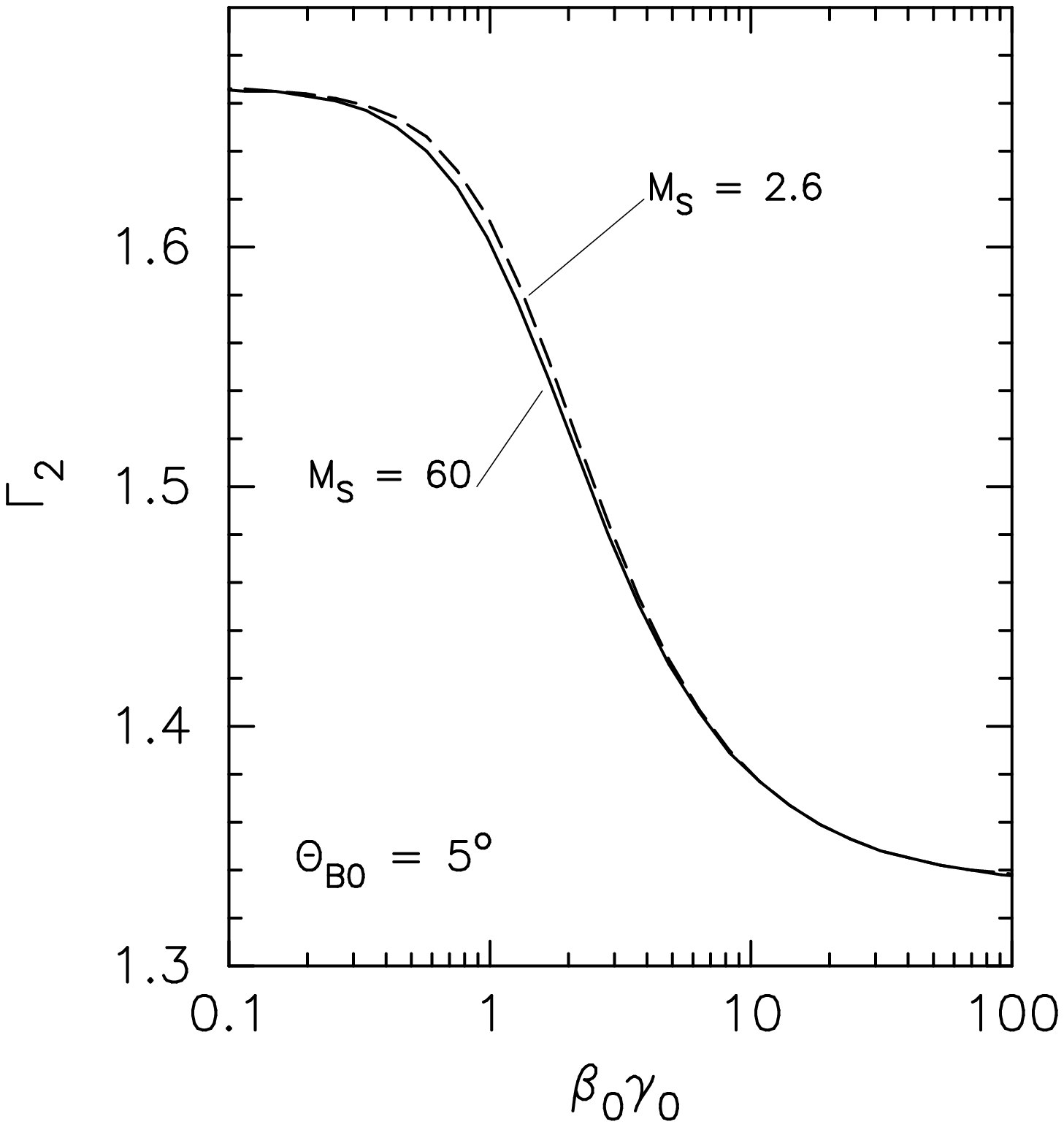} 
\figcaption{The downstream ratio of specific heats, $\Gamma_2$, versus
$\beta_0 \gamma_0$ for the extreme $\MsonicZ$ cases shown in the
left-hand panels of Figure~\ref{fig:T5_6}. Very similar curves are
obtained for all of the examples shown in Figures~\ref{fig:T5_6} and
\ref{fig:T85_6}. 
\label{fig:sp_heats}
}
\end{figure}

In Figure~\ref{fig:sp_heats} we show $\Gamma_2$ for the two extreme
sonic Mach number cases from Figure~\ref{fig:T5_6}. Our approximation
for $\Gamma_2$ depends on $\Vsk$ and $r$, but turns out to be quite
insensitive to $r$, at least in the range above $r \sim 2.7$.  The
$\Gamma_2$ generated by equation~(\ref{SpecHeat}) is essentially the
same as the low temperature solutions presented in Figure~2b of
\citet{HD88} for an $e^+e^-$ plasma.  As noted above, if Fermi
acceleration is permitted to occur, $\Gamma_2$ will approach $4/3$ at
lower $\beta_0 \gamma_0$ due to the contribution of energetic
particles.

The variation of $r$ with shock speed we show here closely resembles
the low sonic Mach number solutions depicted in Figure~4 of
\cite{FK79}, who considered jump conditions in parallel (\iec\ $\MalfZ
\to \infty$) trans-relativistic shocks using an isotropic
J\"uttner-Synge equation of state \citep{Synge57}.  The principal
effect of upstream heating is to lower the compression ratio and
weaken the shock when the ratio of the particle pressure to the rest
mass energy exceeds the square of the shock four-velocity.  The
compression ratio clearly becomes insensitive to the plasma heating in
ultrarelativistic, parallel shocks since $\Gamma_0$ is always $4/3$
and the solution is $\beta_0\beta_2=1/3$ \cite[\egc][]{BM76}.  An
array of possible jump conditions is admitted when an extension to
multi-component plasmas is explored, such as in \citet{Peacock81},
\citet{Kirk88}, \citet{HD88} and \citet{BH91}, where the thermal
interplay of ions and electrons on the shock dynamics in the
trans-relativistic regime can be encapsulated using the two adiabatic
shock index parameters $\Gamma_0$ and $\Gamma_2$.  Such a
parameterization implies that a variety of equations of state can be
accommodated within the formalism presented here.  
Notwithstanding,
when particle acceleration is considered,
thermal equations of state become inappropriate and simulation results
become more important 
\citep{ER91}.
 
\begin{figure}[!hbtp]              
\epsscale{0.65} \plotone{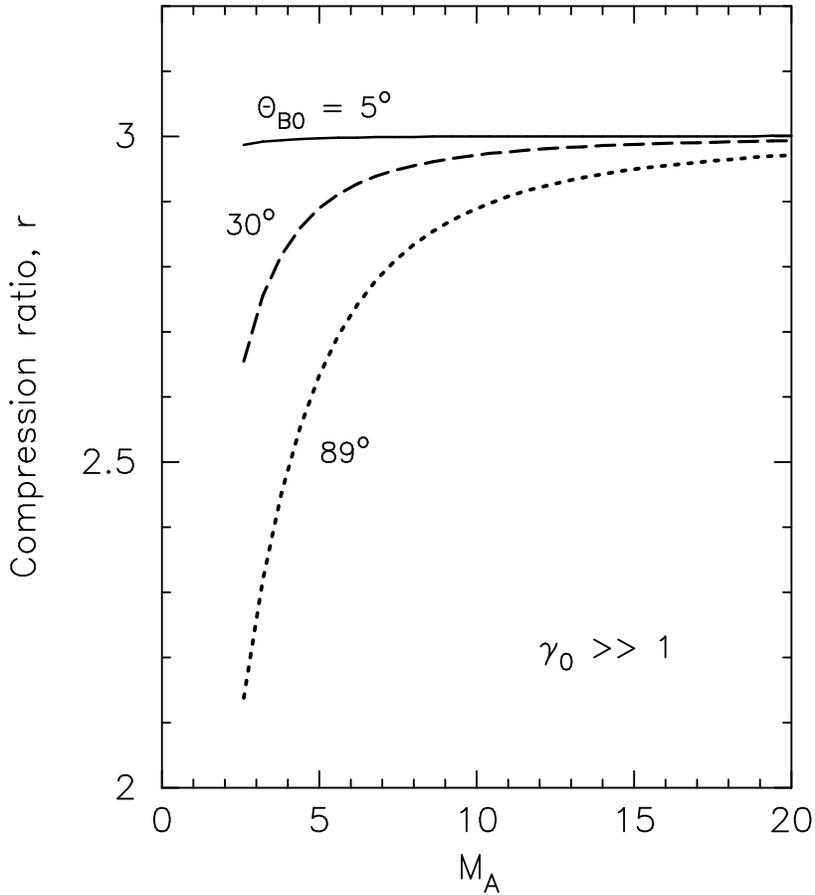} 
\figcaption{Compression ratio versus \alf\ Mach number for various
upstream magnetic obliquities, $\TbnZ$. The values of $r$ are obtained
in the \ultrarel\ limit with $\alpha=1$,
\label{fig:jump_alf}
}
\end{figure}

Figure~\ref{fig:jump_alf} shows how $r$ varies as a function of
$\MalfZ$ in the \ultrarel\ limit for various values of $\TbnZ$ and
with isotropic pressure. The compression ratio depends strongly on the
{\it upstream} obliquity and can drop well below 3 for low $\MalfZ$.
The lower right-hand panels of Figures~\ref{fig:T5_6} and
\ref{fig:T85_6} also show that a low $\MalfZ$ produces magnetic
stresses that cause the downstream flow to deflect from the shock
normal direction, causing $\TuTwo$ to vary inversely as
$\MalfZ$. Despite the fact that $\TuTwo \sim 20^\circ$ for
$\MsonicZ=100$ and $\MalfZ=2.6$, the approximate
equations~(\ref{eq:enflux})-(\ref{eq:pzflux}) give essentially
identical results as the complete equations~(\ref{PE42}-\ref{CM34}).
The impact of either Mach number on $\TBnTwo$ is relatively small.

The most important consequence of a large magnetic field energy
density is that it lowers the compression ratio in oblique shocks (the
field is dynamically passive in parallel ones with $\TbnZ=0$), even
when $\gamma_0\gg 1$.  The effect is contained in Eq.~(4.11) of
\citet{KC84}, which specifies the jump condition or downstream flow
four-velocity for an ultrarelativistic perpendicular MHD shock.
Algebraically simplifying their formula, and expressing it in terms of
the three-velocity compression ratio $r$ and Alfv\'enic Mach number
$\MalfZ$ used here, leads to the form (for $\beta_{x0}=\beta_0\approx
1$)
\begin{equation}
   r\; =\; {{\beta_{0}}\over{\beta_{x2}}}\; \approx\; 
  {{1}\over{2}} \,\biggl\{ \sqrt{\MalfZ^4 + 16 \MalfZ^2 + 16}
        - 2 - \MalfZ^2 \biggr\}
\ ,
 \label{eq:KCjump}
\end{equation}
where the ratio $\sigma$ of the magnetic plus electric energy flux to
the particle energy flux that is used by \citet{KC84} is given
by $\sigma \approx 1/\MalfZ^2 = B_0^2/(4\pi \rho_0 u_0^2)$.    This
result applies to $\gamma_0\gg 1$ regimes, and is reproduced in the
numerical results depicted in the top right hand panel of
Fig~\ref{fig:T85_6} and also the $89^{\circ}$ curve of
Fig~\ref{fig:jump_alf}.

The weakening of \nonrel\ shocks at low \alf\ Mach numbers is widely
understood.
%
Such an effect
is suggested for shocks with $\beta_{x0}\lesssim 0.8$ in \citet{BH91}
and is also somewhat apparent for $\beta_{x0}\lesssim 0.97$ in
\citet{AC88}.  
\citet{KD99} show $r$ as a function of $\Malf$ through the \transrel\
regime for $\TbnZ = 45^\circ$.
The mildly \rel\ regime was appropriate for shocked
jets in active galactic nuclei, the main application of relativistic
MHD shock analyses over a decade ago.  Here, it is evident that this
lowering of $r$ by dynamically important magnetic fields persists up to
arbitrarily high $\gamma_0$, a result obtained by \citet{KC84}, who
applied relativistic MHD to the consideration of perpendicular pulsar
termination shocks.  The more recent association of ultrarelativistic
shocks with cosmic gamma-ray bursts further motivates the extension to
the $\gamma_0\gtrsim 100$ regime.  Accordingly, this magnetic weakening
of the shock has profound implications for the interpretation of
gamma-ray burst spectra and associated emission mechanisms.

The origin of this effect can be attributed to the anisotropic and
intrinsically relativistic nature of the field structure.  In the
\ultrarel\ regime, the equation of state of a quasi-isotropic,
turbulent field structure replicates that of the familiar
ultrarelativistic gas with $\Gamma=4/3$.  However, if the field is
laminar and oblique, the stress-energy tensor in
equation~(\ref{EMT11}) exhibits anisotropic stresses.  This anisotropy
influences the flux conservation relations for dynamically important
fields when $B_{z0}\neq 0$; moreover it should mimic to some extent
the behavior anticipated from anisotropies due to the gas
contribution.  Such a parallel obviously emerges from results
presented in the next Section.

\begin{figure}[!hbtp]              
\epsscale{0.6} \plotone{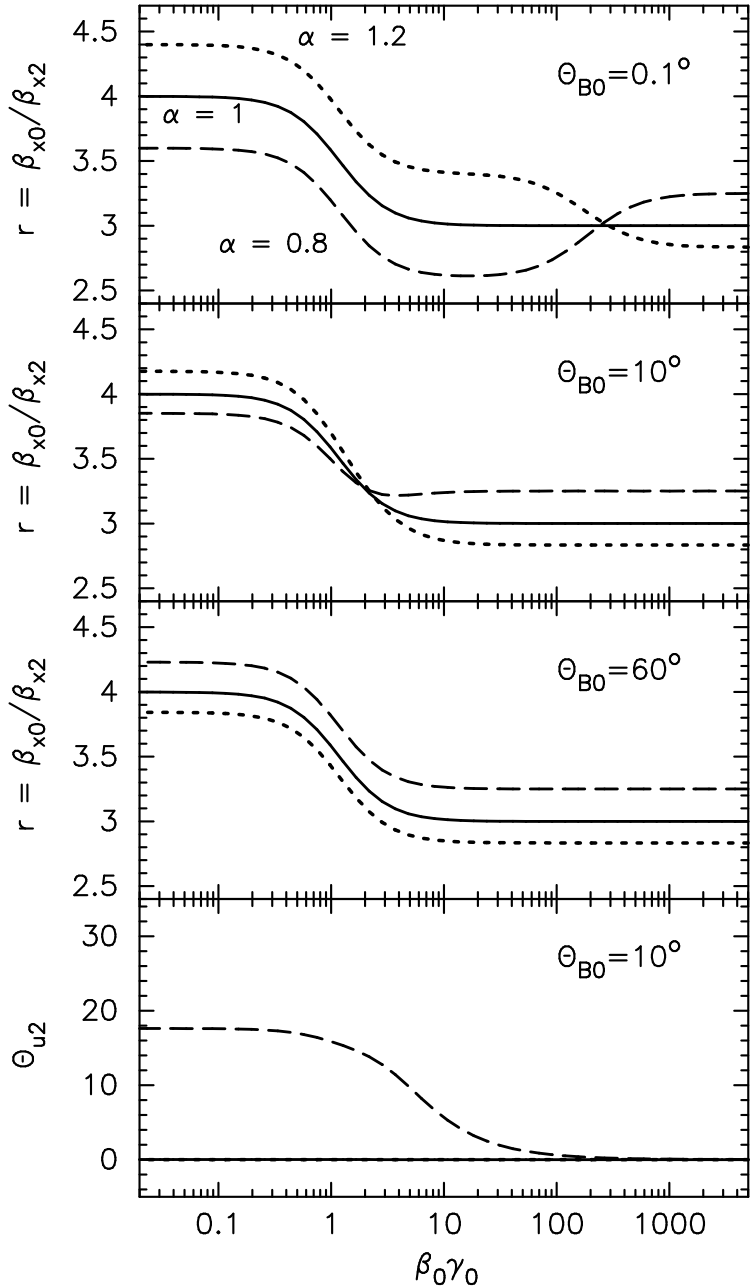} 
\figcaption{The top three panels show the compression ratio, $r$,
versus $\beta_0 \gamma_0$ for shocks with varying downstream
anisotropy, $\alpha=P_{\perp}/P_{\parallel}$, and obliquity, $\TBnZ$.
In all panels, the solid curves have $\alpha=1$, the dashed curves
have $\alpha=0.8$, and the dotted curves have $\alpha=1.2$
downstream. The upstream plasma is taken to be isotropic (\iec\
$\alpha=1$) in all cases, but the upstream $\alpha$ has virtually no
influence on the solutions for the parameter regime we consider. The
anisotropy has little effect on $\TbnTwo$, but does cause
modifications in $\TuTwo$ at mildly \rel\ shock speeds, as indicated
in the bottom panel which shows the angle the downstream flow makes
with the shock normal for the particular case $\TbnZ=10^\circ$.  These
are high Mach number examples ($\MsonicZ = \MalfZ \gtrsim 100$); the
effect on $r$ at lower Mach numbers is shown in
Figure~\ref{fig:r_vs_ani}.
\label{fig:aniso}
}
\end{figure}

\subsection{Gyrotropic Pressure}
 \label{sec:gyrotropicP}
For gyrotropic pressure, an additional constraint is needed to close
the system of equations and obtain an analytical solution; we impose
this via
\begin{equation}
\label{PressConst}
\Pperp = \alpha \Ppar \ ,
\end{equation}
where $\alpha$ is an arbitrary parameter and $\alpha = 1$ corresponds
to isotropic pressure.  Equation~(\ref{PressConst}) allows us to
illustrate the effects of anisotropic pressure, but is not suggested as
a model for specific acceleration scenarios.  In relativistic shocks
where the accelerated population contributes significantly to the total
dynamical pressure, the value of $\alpha$ should deviate significantly
from unity in both the upstream and downstream regions.  Using
equation~(\ref{FTrot}), equation~(\ref{PressConst}) then yields
\begin{equation}
P_{xx} =
\Ppar (\cos^2{\TBn} + \alpha \sin^2{\TBn})
\ ,
\end{equation}
\begin{equation}
P_{xz} =
\Ppar (\alpha - 1) \sin{\TBn} \cos{\TBn}
\ ,
\end{equation}
and using the fluxes given in equations~(\ref{PE42})--(\ref{CM34}) or
(\ref{eq:enflux})--(\ref{eq:pzflux}), we have a closed set of
equations for the jump conditions for shocks with gyrotropic pressure,
arbitrary obliquity, and arbitrary flow speed. Results for various
$\TbnZ$'s and $\alpha$'s are shown in Figure~\ref{fig:aniso} (these
results all have $\MsonicZ = \MalfZ = 100$ but they remain unchanged
for larger Mach numbers). The solid curves have $\alpha=1$, the dashed
curves have $\alpha=0.8$, and the dotted curves have $\alpha=1.2$.  In
all cases, we have taken the pressure in the unshocked gas to be
isotropic and $\alpha \ne 1$ is only applied downstream.

While our solutions can allow for anisotropic upstream pressure, an
upstream $\alpha \ne 1$ generally produces insignificant changes to our
results for the parameter regime discussed here.  Notable exceptions
arise when the sonic Mach number is very low, i.e.
$\gamma_0\MsonicZ^2\sim 1$.  We observe that angular distributions
at relativistic shocks generated by Monte Carlo simulations
\citep[\egc][]{BedOstrow98,ED2002} and semi-analytic
convection-diffusion equation solutions \citep[][]{KGGA2000} indicate
that for plane-parallel scenarios with $\TBnZ=0$ the accelerated
particles are dominated by parallel pressure upstream ($\alpha <1$) and
perpendicular pressure downstream ($\alpha >1$) when near the shock.
This would suggest a possible weakening of the shock if the non-thermal
population were to contribute significantly to the dynamics, thereby
rendering such a contribution less likely.  The picture may be 
much different for oblique and perpendicular shocks.

\begin{figure}[!hbtp]              
\epsscale{0.6} \plotone{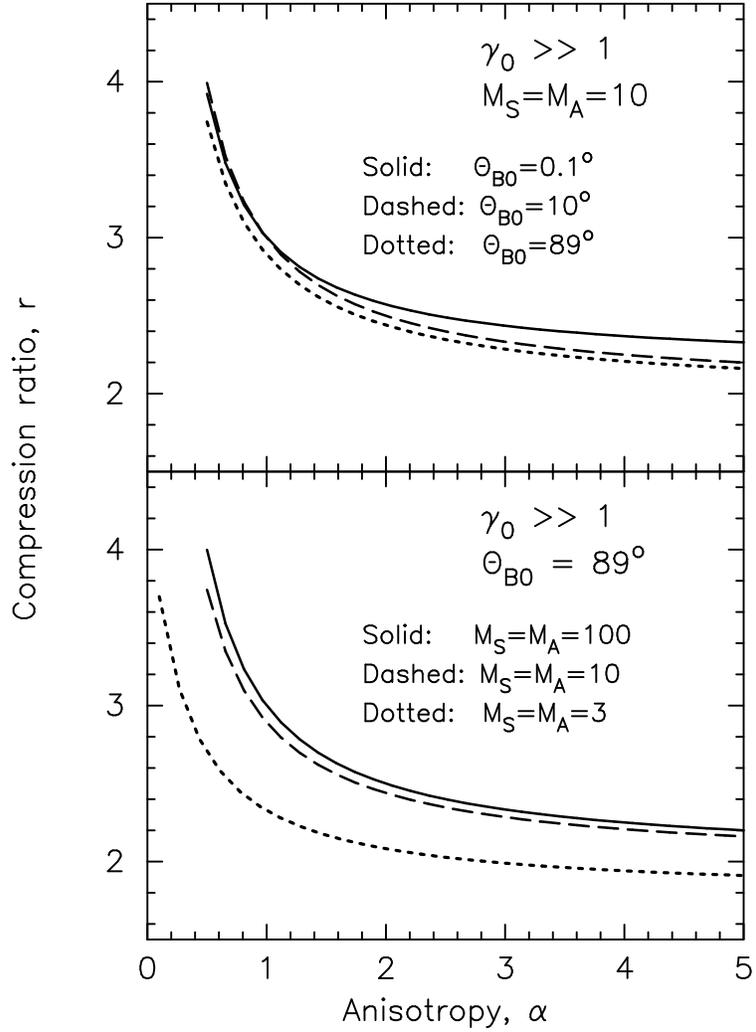} 
\figcaption{Compression ratio versus anisotropy for \ultrarel\
shocks. The top panel shows the effects of $\TbnZ$, while the bottom
panel shows the effects of Mach number keeping $\TbnZ=89^\circ$. Note
that, depending on $\alpha = P_{\perp}/P_{\parallel}$, the
compression ratio $r$ can be either larger or smaller than 3.
\label{fig:r_vs_ani}
}
\end{figure}

The effects of anisotropic pressure on the compression ratio depend
strongly on $\gamma_0$ and $\TbnZ$. 
When $\TbnZ$ is small, the downstream angle $\TbnTwo$ is also small at
\nonrel\ and mildly \rel\ shock speeds (top panel of
Figure~\ref{fig:aniso}).  Therefore, at these speeds $P_{xz} \sim 0$
and $P_{xx} \sim \Ppar$ for any $\alpha$.  Since $\Ppar =
\Pperp/\alpha$, the fraction of downstream pressure in $P_{xx}$ is
inversely proportional to $\alpha$ and since $P_{xx}$ largely
determines $r$, the compression ratio is less than the isotropic
value, $\rIso$, for $\alpha= 0.8$ and greater than $\rIso$ for
$\alpha=1.2$, as shown in the figure.
In contrast, as $\gamma_0 \to \infty$, $\TBnTwo$ approaches $90^\circ$
for any $\TBnZ>0$ (see Figure~\ref{fig:T5_6}) and $P_{xx} \sim \Pperp$
with $P_{xz}$ again approximately equal to zero.  Now, the fraction of
pressure in $P_{xx}$ will be proportional to $\alpha$ and $r < \rIso$
for $\alpha >1$ and $r > \rIso$ for $\alpha <1$.
The transition where $r$ crosses $\rIso$ occurs at slower shock speeds
as $\TBnZ$ increases (see the $\TbnZ=10^\circ$ panel of
Figure~\ref{fig:aniso}) until $\TBnZ$ is large enough
($\TbnZ=60^\circ$ panel) so no transition occurs. While the examples
in Figure~\ref{fig:aniso} all have $\MsonicZ = \MalfZ = 100$, the
transition $\beta_0 \gamma_0$ is independent of Mach number.
The dependence of $\TBnTwo$ on $\alpha$ is relatively small for the
examples shown in Figure~\ref{fig:aniso} but $\TuTwo$ can change
significantly with anisotropy, as shown in the bottom panel for the
$\TBnTwo =20^\circ$ example.

In Figure~\ref{fig:r_vs_ani} we show $r$ as a function of $\alpha$ for
various Mach numbers and $\TbnZ$'s, all in the \ultrarel\ limit. The
top panel shows that $r$ is relatively insensitive to the upstream
magnetic field angle, with the lower panel showing a somewhat greater
sensitivity to Mach number. The most important aspect of these plots
is the fact that $r$ can be either higher or lower than the canonical
value of 3 if anisotropic pressure is important. This contrasts with
the effects of a magnetic field (with isotropic pressure), where low
$\Malf$'s in the \ultrarel\ limit gave $r < 3$ only
(Figures~\ref{fig:T85_6} and \ref{fig:jump_alf}).  

\subsubsection{Analytic solution for $\gamma_0 \gg 1$}
Using our flux conservation relations in the limit $\gamma_0 \gg
1$, and assuming that the upstream pressure is isotropic and the
downstream adiabatic index $\Gamma_2
= 4/3$, we obtain an analytic solution for $r$ valid for any oblique
angle and any $\MsonicZ$ or $\MalfZ$ greater than one:
\begin{equation}
\label{eq:r_aniso_cube}
(\alpha +1) b_0 r^3 + 2\alpha (w_0 + b_0)r^2 -
\left[2(3\alpha + 1)w_0 + (7\alpha + 3) b_0\right ] r +
2(2\alpha + 1)(w_0 + b_0) = 0
\ ,
\end{equation}
where $w_0 = e_0 + P_0$ is the enthalpy density and $b_0 \equiv
B_{0z}^2/4\pi$.
%
%
%
Only the
$z$-component of $\hbox{\bf B}$ appears since $B_{x2} = B_{x0}$ and
this component drops out.
Eliminating the $r=1$ root gives
\begin{equation}
\label{eq:r_aniso_quad}
(\alpha + 1) b_0 r^2 + \left[2\alpha w_0 + (3\alpha + 1) b_0\right]r -
2(2\alpha + 1)(w_0 + b_0)\ =0
\ ,
\end{equation}
or,
\begin{equation}
\label{eq:r_anis_sol}
r = \bfrac{\left\{\left[2\alpha q + (3\alpha + 1)\right]^2 + 
8(\alpha + 1)(2\alpha + 1)(q + 1)\right\}^{1/2} -
2\alpha q - (3\alpha + 1)}
{2(\alpha + 1)}
\ ,
\end{equation}
where $q \equiv w_0/b_0$.
Solutions in terms of $\Malf$ and $\TbnZ$ can be found using the
relation $ b_0= \rho_0 c^2 \sin^2{\TbnZ}/\Malf^2$.
For isotropic pressure, i.e., $\alpha=1$, 
\begin{equation}
\label{eq:r_approx}
r = 
\left [  \left (\frac{q}{2} + 1\right )^2 + 3 (q +1) \right ]^{1/2}
- \left (\frac{q}{2} + 1 \right )
\ .
\end{equation}
Equations~(\ref{eq:r_approx}) and (\ref{eq:r_anis_sol}) reproduce
the results
obtained with the exact equations shown in Figures~\ref{fig:jump_alf}
and \ref{fig:r_vs_ani} (bottom panel, solid curve)
to a high degree of accuracy.
In the limit of $\MsonicZ \gg 1$ and $\TbnZ \too 90^\circ$,
equation~(\ref{eq:r_approx}) is
equivalent to equation~(\ref{eq:KCjump}) from \citet{KC84}.
Once $r$ is determined, the downstream $z$-component of $\hbox{\bf B}$
can be found from equation~(\ref{eq:DSmag}).
%

%
%
%
%
%
%

\section{CONCLUSIONS}
We have derived shock jump conditions for arbitrary shock speeds and
obliquities. When combined with a simple approximation for the ratio
of specific heats (equation~\ref{SpecHeat}), these equations specify
the downstream conditions in terms of upstream parameters for
isotropic pressure. For the case of gyrotropic pressure, an additional
arbitrary parameter (equation~\ref{PressConst}) is required to close
the set of equations and we have presented a number of solutions where
the downstream pressure is gyrotropic.  The exact equations for
energy and momentum conservation are fairly complicated, but we have
presented simpler, approximate results
(equations~\ref{eq:enflux}-\ref{eq:pzflux}) which are extremely
accurate in a wide parameter regime (\iec\ $\MsonicZ > 5$; $\MalfZ >
5$; $0.8 < \alpha < 1.5$) for any shock speed or obliquity.  To our
knowledge, this is the first presentation of oblique shock solutions
with gyrotropic pressure that continuously span the domains of
\nonrel, \transrel, and \ultrarel\ shocks of arbitrary obliquity.
In addition, we have presented an analytic solution for the shock
compression ratio, $r$, in the limit of \ultrarel\ shock speeds valid
for oblique shocks of any $\MsonicZ$ or $\MalfZ$ with gyrotropic pressure
(equation~\ref{eq:r_anis_sol}).

The results presented here assume that no first-order Fermi
acceleration occurs, but they apply directly to test-particle
acceleration where the energy density in accelerated particles is
small. They also constitute an important ingredient in more complex
models of nonlinear particle acceleration.  In the test-particle case,
the compression ratio $r$ is altered by large magnetic fields and/or
anisotropic pressures, even at \ultrarel\ speeds.  The observation of
marked departures from the canonical value of $r=3$ for the
compression ratio of an ultrarelativistic shock, for either
dynamically important fields or significant pressure anisotropies, is
a major result of this paper.  In the case of magnetic field
influences on the dynamics of relativistic shocks, our results extend
the conclusions of \citet{KC84} to all shock obliquities.  The
similarity of such consequences of fluid anisotropy and low Alfv\'enic
Mach numbers is a consequence of the similar nature of the plasma and
electromagnetic contributions to the spatial components of the
stress-energy tensor.

The importance of this result is obvious, since changes in $r$ map
directly to changes in the power-law index of the accelerated
spectrum.  This power law is the most important characteristic of
test-particle Fermi acceleration, and is usually associated with
\transrel\ internal shocks in gamma-ray burst (GRB) models
\citep[\egc][]{RM92,Piran99}.  In addition, the outer blast wave is an
ultrarelativistic shock during most of its active phase, sweeping up
and accelerating interstellar material. This shock, believed to
produce long-lasting afterglows, eventually transitions to a
non-relativistic phase.  Our results are directly applicable to
test-particle acceleration models of both the internal and external
shocks in GRBs, as well as to shocks believed to exist in jets in
active galactic nuclei.  In future work, we will apply these jump
conditions to nonlinear shock acceleration models where the
accelerated particle population can modify the shock structure.

\newcommand{\aaDE}[3]{ 19#1, A\&A, #2, #3}
\newcommand{\aatwoDE}[3]{ 20#1, A\&A, #2, #3}
\newcommand{\aasupDE}[3]{ 19#1, {\itt A\&AS,} {\bff #2}, #3}
\newcommand{\ajDE}[3]{ 19#1, {\itt AJ,} {\bff #2}, #3}
\newcommand{\anngeophysDE}[3]{ 19#1, {\itt Ann. Geophys.,} {\bff #2}, #3}
\newcommand{\anngeophysicDE}[3]{ 19#1, {\itt Ann. Geophysicae,} {\bff #2}, #3}
\newcommand{\annrevDE}[3]{ 19#1, {\itt Ann. Rev. Astr. Ap.,} {\bff #2}, #3}
\newcommand{\apjDE}[3]{ 19#1, {\itt ApJ,} {\bff #2}, #3}
\newcommand{\apjtwoDE}[3]{ 20#1, {\itt ApJ,} {\bff #2}, #3}
\newcommand{\apjletDE}[3]{ 19#1, {\itt ApJ,} {\bff  #2}, #3}
\newcommand{\apjlettwoDE}[3]{ 20#1, {\itt ApJ,} {\bff  #2}, #3}
\newcommand{\apjpress}{{\itt ApJ,} in press}
\newcommand{\apjletpress}{{\itt ApJ(Letts),} in press}
\newcommand{\apjsDE}[3]{ 19#1, {\itt ApJS,} {\bff #2}, #3}
\newcommand{\apjsubDE}[1]{ 19#1, {\itt ApJ}, submitted.}
\newcommand{\apjsubtwoDE}[1]{ 20#1, {\itt ApJ}, submitted.}
\newcommand{\appDE}[3]{ 19#1, {\itt Astroparticle Phys.,} {\bff #2}, #3}
\newcommand{\apptwoDE}[3]{ 20#1, {\itt Astroparticle Phys.,} {\bff #2}, #3}
\newcommand{\apppressDE}[1]{ 20#1, {\itt Astroparticle Phys.,} in press}
\newcommand{\araaDE}[3]{ 19#1, {\itt ARA\&A,} {\bff #2},
   #3}
\newcommand{\assDE}[3]{ 19#1, {\itt Astr. Sp. Sci.,} {\bff #2}, #3}
\newcommand{\icrcplovdiv}[2]{ 1977, in {\itt Proc. 15th ICRC(Plovdiv)},
   {\bff #1}, #2}
\newcommand{\icrcsaltlake}[2]{ 1999, {\itt Proc. 26th Int. Cosmic Ray Conf.
    (Salt Lake City),} {\bff #1}, #2}
\newcommand{\icrcsaltlakepress}[2]{ 19#1, {\itt Proc. 26th Int. Cosmic Ray Conf.
    (Salt Lake City),} paper #2}
\newcommand{\jgrDE}[3]{ 19#1, {\itt J.G.R., } {\bff #2}, #3}
\newcommand{\jppDE}[3]{ 19#1, {\it J. Plasma Phys., } {\bf #2}, #3}
\newcommand{\mnrasDE}[3]{ 19#1, {\itt M.N.R.A.S.,} {\bff #2}, #3}
\newcommand{\mnrastwoDE}[3]{ 20#1, {\itt M.N.R.A.S.,} {\bff #2}, #3}
\newcommand{\mnraspress}[1]{ 20#1, {\itt M.N.R.A.S.,} in press}
\newcommand{\natureDE}[3]{ 19#1, {\itt Nature,} {\bff #2}, #3}
\newcommand{\pfDE}[3]{ 19#1, {\itt Phys. Fluids,} {\bff #2}, #3}
\newcommand{\phyreptsDE}[3]{ 19#1, {\itt Phys. Repts.,} {\bff #2}, #3}
\newcommand{\physrevEDE}[3]{ 19#1, {\it Phys. Rev. E,} {\bf #2}, #3}
\newcommand\prDE[3]{ 19#1, {\it Phys. Rev.,} {\bf #2}, #3}
\newcommand{\prlDE}[3]{ 19#1, {\it Phys. Rev. Letts,} {\bf #2}, #3}
\newcommand{\revgeospphyDE}[3]{ 19#1, {\itt Rev. Geophys and Sp. Phys.,}
   {\bff #2}, #3}
\newcommand{\rppDE}[3]{ 19#1, {\itt Rep. Prog. Phys.,} {\bff #2}, #3}
\newcommand{\ssrDE}[3]{ 19#1, {\itt Space Sci. Rev.,} {\bff #2}, #3}

\end{document}